\documentclass[smallcondensed,natbib]{svjour3} 
\smartqed  
\usepackage{graphicx}
\usepackage{mathptmx}      
\usepackage{latexsym}
\begin{document}

\title{Where are the Uranus Trojans?}

\subtitle{}


\author{
        R. Dvorak \and {\'A}. Bazs{\'o} \and L.-Y. Zhou
}


\institute{
        R. Dvorak \and {\'A}. Bazs{\'o}\\
        Institute of Astronomy, University of Vienna,
        T{\"u}rkenschanzstr. 17, A-1180 Wien, Austria\\
        \email{rudolf.dvorak@univie.ac.at}\\
        \at
        L.-Y. Zhou\\
        Department of Astronomy, Nanjing University,
        Nanjing 210093, China\\
}

\date{Received: date / Accepted: date}

\maketitle

\begin{abstract}
The area of stable motion for fictitious Trojan asteroids around Uranus'
equi\-la\-te\-ral equilibrium points is investigated with respect to the
inclination of the asteroid's orbit to determine the size of the regions and
their shape.
For this task we used the results of extensive numerical integrations of orbits
for a grid of initial conditions around the points $L_4$ and $L_5$, and
analyzed the stability of the individual orbits.
Our basic dynamical model was the Outer Solar System (Jupiter, Saturn, Uranus
and Neptune).
We integrated the equations of motion of fictitious Trojans in the vicinity of
the stable equilibrium points for selected orbits up to the age of the
Solar system of $5 \times 10^{9}$ years.
One experiment has been undertaken for cuts through the Lagrange points for
fixed values of the inclinations, while the semimajor axes were varied.
The extension of the stable region with respect to the initial semimajor
axis lies between $19.05 \le a \le 19.3$ AU but depends on the initial
inclination.
In another run the inclination of the asteroids' orbit was varied in the range
$0^\circ < i < 60^\circ$ and the semimajor axes were fixed.
It turned out that only four 'windows' of stable orbits survive: these are the
orbits for the initial inclinations $0^\circ < i < 7^\circ$, $9^\circ < i <
13^\circ$, $31^\circ < i < 36^\circ$ and  $38^\circ < i < 50^\circ$.
We postulate the existence of at least some Trojans around the Uranus
Lagrange points for the stability window at small and also high inclinations.


\end{abstract}

\section{Introduction}

The first discovery of a Jupiter Trojan in 1906 (Achilles by Max Wolf in
Heidelberg) proved that the equilateral equilibrium points in a simplified
dynamical model Sun -- planet -- massless body are not only of hypothetical
interest. Ever since many of such Trojan asteroids of Jupiter have been found
and now we have knowledge of several thousands of objects in the 1:1 mean motion
resonance with Jupiter. Several investigations show the symmetry of these two
stable equilibrium points not only in the restricted three body problem
(\citet{San2002, Erd1988}) but also in the realistic dynamical model of the
Outer Solar System (OSS) consisting of Jupiter, Saturn, Uranus and Neptune, like
in e.g. \citet{Dvo2005, Fre2006, Sch2004, Nes2000, Rob2005, Tsi2005a}.

But we have also evidence for the existence of asteroids in orbit around the
equilateral equilibrium points of Neptune and also of Mars. Since the
discovery of the first Neptune Trojan (\citet{Chi2003}) -- the minor planet 2001
QR322 around its equilibrium point $L_5$ -- a lot of work has been dedicated to
understand the dynamics of Neptune Trojans. On one hand the stability of
hypothetical bodies was studied in the model of the Outer Solar System, on the
other hand long term investigations tried to understand how Trojans can survive,
or even get captured in these kind of orbits when one allows the large planets
to migrate. An extensive stability study of the Trojans for the four gas
giants has been undertaken by \citet{Nes2002} in different dynamical models.
The results of the realistic n-body simulations for clouds of hypothetical
Trojans around the Lagrange point $L_4$ showed how fast the depletion of
asteroids is acting for Saturn, Uranus and Neptune. For several hundreds of
bodies, different eccentricities and inclinations as well as semimajor axes the
integrations were carried out over 4 Gyrs. Saturn Trojans turned out to be
unstable already after several $10^5$ years, Uranus Trojans were also found to
be unstable but only after several million years, whereas Neptune's Trojans
mostly survived for the whole integration time of 4 Gyrs. In a similar study for
Uranus and Neptune Trojans \citet{Mar2003} determined the diffusion speed of
Trojans and derived even expressions for the secular frequencies using their
numerical results.
Both studies were undertaken for initial inclinations of the Trojans up to
$i=30^{\circ}$ and showed quite similar results for the dependence on the
inclination.

Taking into account the early evolution of the Solar System the migration
processes need to be included. This has been done in a paper by \citet{Kor2004}
for the Neptune Trojans where it has been shown that only $5\%$ of an initial
Trojan population could survive this early stage when Jupiter migrated inward
(5.4 to 5.2 AU), and Saturn (8.7 to 9.5 AU), Uranus (16.2 to 19.2 AU) and
Neptune (23 to 29 AU) migrated outwards. Three different theories for the origin
of Neptune Trojans are developed and tested by \citet{Chi2005}, they use the
results to constrain the circumstances of the formation of planets and the
accretion of Trojans. In a more recent study \citet{Nes2009} in the NICE model
(e.g. \citet{Tsi2005b}) -- where Uranus and Neptune could interchange their
location in the Solar System and Jupiter and Saturn go through the 2:1 Mean Motion
Resonance -- it was shown that the Neptune Trojans were captured by a process
quite similar to the chaotic capture of Jupiter Trojans (\citet{Mor2005}).
A similar process can be assumed for the Uranus Trojans, but up to now no Uranus
Trojans have been detected which can be explained by the results of the pure
dynamical studies mentioned before (\citet{Nes2002, Mar2003}).
These investigations stopped at an inclination of $i=30^{\circ}$; the discovery
of a highly-inclined Neptune Trojan (\citet{She2006}) raised the question of the
general stability of Trojans at high inclinations. In a recent study concerning
the Neptune Trojans (\citet{Zho2009}) it was shown that especially for larger
inclinations Trojans may be in stable orbits. We have therefore undertaken
an extensive numerical study of the -- hypothetical -- Uranus Trojan cloud
for the whole range of inclinations $0^{\circ} \le i \le 60^{\circ}$.

\section{The method of investigation}

\subsection{The model and the grid of initial conditions}

As dynamical model for the determination of the stable regions around Uranus'
$L_4$ and $L_5$ points the Outer Solar System was considered,
consisting of the Sun and the four gas giants. The fictitious asteroids were
assumed to be massless. No test computations had to be undertaken, because former
studies for the Jupiter and the Neptune Trojans provided
realistic results for that model (\citet{Dvo2008, Zho2009, Erd2009}).
Thus the equations of motion (only in the Newtonian framework) were integrated
using the Lie-series method, a code which uses an adaptive step size and which
has already been extensively tested and used in many numerical studies
(e.g. \citet{Han1984, Lic1984, Del1984}).

Different time scales were tested: First estimations have been undertaken for
only $10^6$ years of integration time to separate the regions
in phase space which are more or less immediately unstable.
Then, for the remaining stable orbits, the integrations were extended up to $10^7$
and $10^8$ years, and finally for three selected orbits the time scale was
set to the age of the Solar system ($5 \times 10^9$ years).

The initial conditions in the vicinity of the triangular equilibrium points
were chosen such that the orbital elements mean anomaly,
eccentricity and longitude of the node for the Trojans were set to the corresponding
values of Uranus and only the argument of the perihelion was set to $\omega =
\omega_{ Uranus} \pm 60^{\circ}$.

For one run different values of the semimajor axes close to the one of Uranus
were taken: $a_{Trojan}=a_{Uranus} \pm m \times 0.007$ AU, for $1 \leq m \leq 50$.
The inclination was set to be equal to the one of Uranus. For a second run the semimajor
axis was set to the one of Uranus, but now the inclinations of the fictitious
Trojans were changed in the range $i_{Uranus} \leq i_{Trojan} \leq 60^\circ$
for steps of $\Delta i = 1^\circ$.  Both studies were undertaken separately
for the leading and the trailing Lagrange-points (see chapter 4).
For the whole grid of initial conditions around $L_4$ we integrated some
thousand orbits to get a first picture of the stable region (see chapter 3,
Fig. \ref{fig1}).

\subsection{The methods of analysis}

Different tools of analysis were used to determine on one hand the stable regions in
phase space and on the other hand to find out the reason for the instability of orbits:

\begin{itemize}

\item the maximum eccentricity within the integration time,

\item the libration amplitude around the Lagrange point,

\item the escape time for the unstable orbits,

\item the involved frequencies.

\end{itemize}

For the first three tools there is not much to say about, because they are self
explaining, except that we define the libration width (or libration
amplitude) as the difference of the maximum and minimum value of the mean
longitude-difference of the Trojan with respect to Uranus.
The program package {\sc SigSpec} used for the frequency analysis was provided
by \citet{Ree2007}. This is a code which is very efficient when used for data
with equidistant time-domain sampling intervals\footnote{This does NOT mean that
we used a constant step size for the numerical integration of the equations of motion.}, but also
for non-equally spaced data. The computer program is a novel
method in time series analysis\footnote{It is the first technique relying on an
analytic solution for the
probability distribution produced by white noise in an amplitude
spectrum (Discrete Fourier Transform). Returning the probability that
an amplitude level is due to a random process, it provides an
objective and unbiased estimator for the significance of a detected
signal.}. It incorporates the frequency and phase of the
signal appropriately and takes also into account the properties of the
time-domain sampling. Thus it provides a very accurate peak
detection (\citet{Kal2008}). The benefit of frequency- and phase-resolved statistics and the capability
of full automation are important also in this application with
equidistant data.

\section{Stability of $L_4$ depending on the inclination: a first overview}

\begin{figure}
\begin{center}
\includegraphics[width=5.in,angle=0]{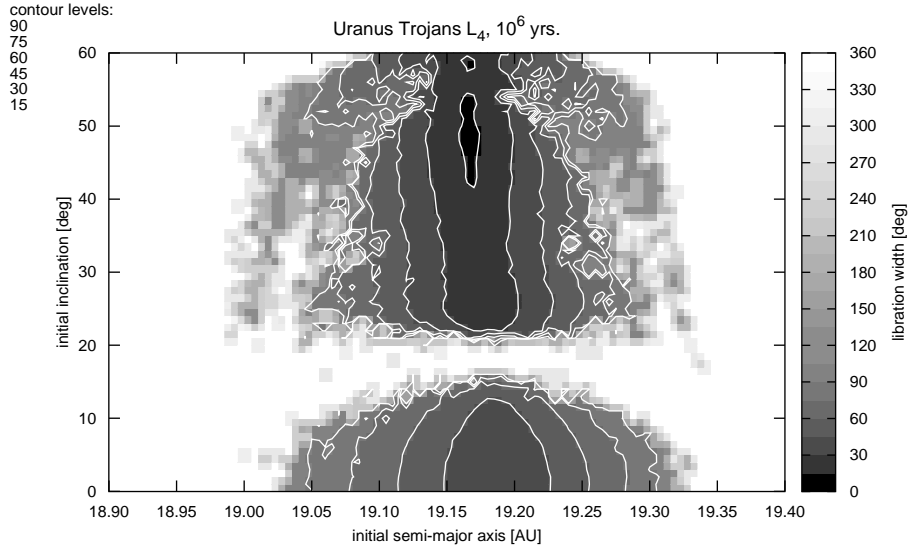}
\caption{
  The initial semi-major axes of the fictitious object (x-axis) are
  plotted versus the initial inclination of the Trojan (y-axis). The grey-tone
  visualizes the libration amplitude around the Lagrange point $L_4$; the contour-lines
  show the limits of the libration amplitudes with a step of $15^\circ$ (for
  detail see in the text)}
\label{fig1}
\end{center}
\end{figure}

To get an idea of the stability we used the results of numerically integrated
orbits of several thousand fictitious Trojans for an integration time of
$10^6$ years. For an increment of
$1^\circ$ for inclinations between $0^\circ \le i_{Trojan} \le 60^\circ$ we used 100
orbits with different semimajor axes defined before. As mentioned we used
three different methods for the
analysis, which provided all more or less the same results concerning the
stability of an orbit. In Fig. \ref{fig1} we mark the amplitudes $A$
of libration around the $L_4$ point in different greytones and with contour-lines. The
innermost lines on the upper part of the graph ($i \ge 20^\circ$) enclose
orbits in the stable region with $A \le 15^\circ$ (black region); the fuzzy edges show the coexistence of stable and unstable orbits in the transition regions.
Furthermore it is interesting to note that for small inclinations there
are no orbits with amplitudes of libration smaller than $A \le 30^\circ$ present,
which was already mentioned in the paper by \citet{Mar2003}. On the contrary
quite small libration amplitudes occur for large inclination ($42^{\circ} \le
i \le 52^{\circ}$). The non existence of orbits from moderate inclinations on
($15^{\circ} \le i \le 30^{\circ}$) was already found in the papers by
\citet{Nes2000, Nes2002} and \citet{Mar2003}; in our study we can confirm it,
but then, for larger eccentricities, new regions of stable motion
appear\footnote{compare section 4.2, Fig.\ref{fig4}.}.

\section{The Results of the Cuts}

The integration of thousands of orbits for longer time would need a huge
amount of computer time -- even with a fast program and several CPUs for
running the code. We decided to limit the integration time to $10^7$ years (or
$10^8$ years where appropriate) in the dynamical model of the OSS and specified
the following initial conditions for the fictitious Trojans: the initial orbital
elements for the massless asteroids were set to the one of Uranus, but for the
$L_4$ Trojans we set $\omega = \omega + 60^\circ$ and for the $L_5$ Trojans
$\omega = \omega - 60^\circ$. To be able to determine the sizes of stable regions
with respect to the semimajor axes and the inclination of the Trojans we
chose the following two different experiments:

\begin{itemize}
\item {\bf a-cut} where we varied the initial semimajor axes of
100 fictitious Trojans and set the initial inclination -- as an illustrative example -- to
$i = 3^\circ$; the integration time was limited to $10^7$ years.

\item {\bf i-cut} where we fixed the semimajor axis to the one of Uranus and set the inclination
of 60 fictitious Trojans to $1^\circ < i < 60^\circ$; the integration time was
set to $10^8$ years.

\end{itemize}

In addition two different runs -- independently for $L_4$ and $L_5$ --
have been undertaken to check any possible differences.

\subsection{The {\bf a-cuts}}

\begin{figure}
\begin{center}
\includegraphics[width=4.in,angle=0]{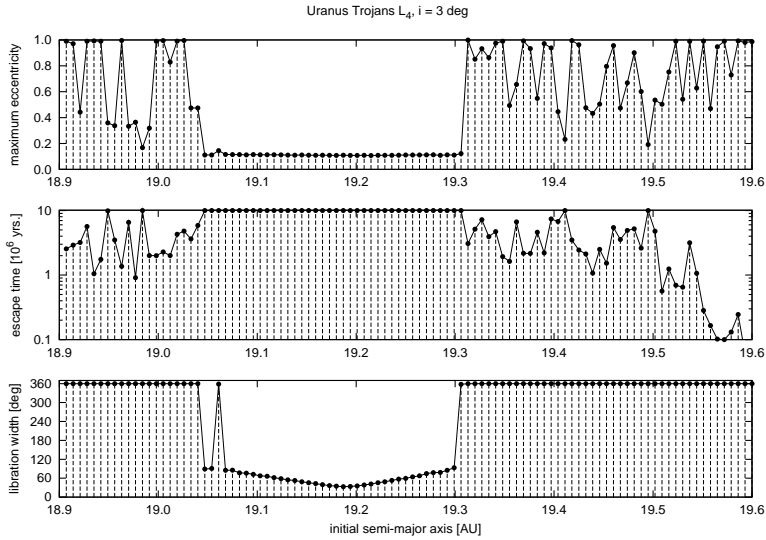}
\caption{{\bf a-cut} for $L_4$: semimajor axis versus the maximum eccentricity
  (top), versus the escape time (middle) and versus the libration width around the
  Lagrange point (bottom).}
\label{fig2}
\end{center}
\end{figure}

\begin{figure}
\begin{center}
\includegraphics[width=4.in,angle=0]{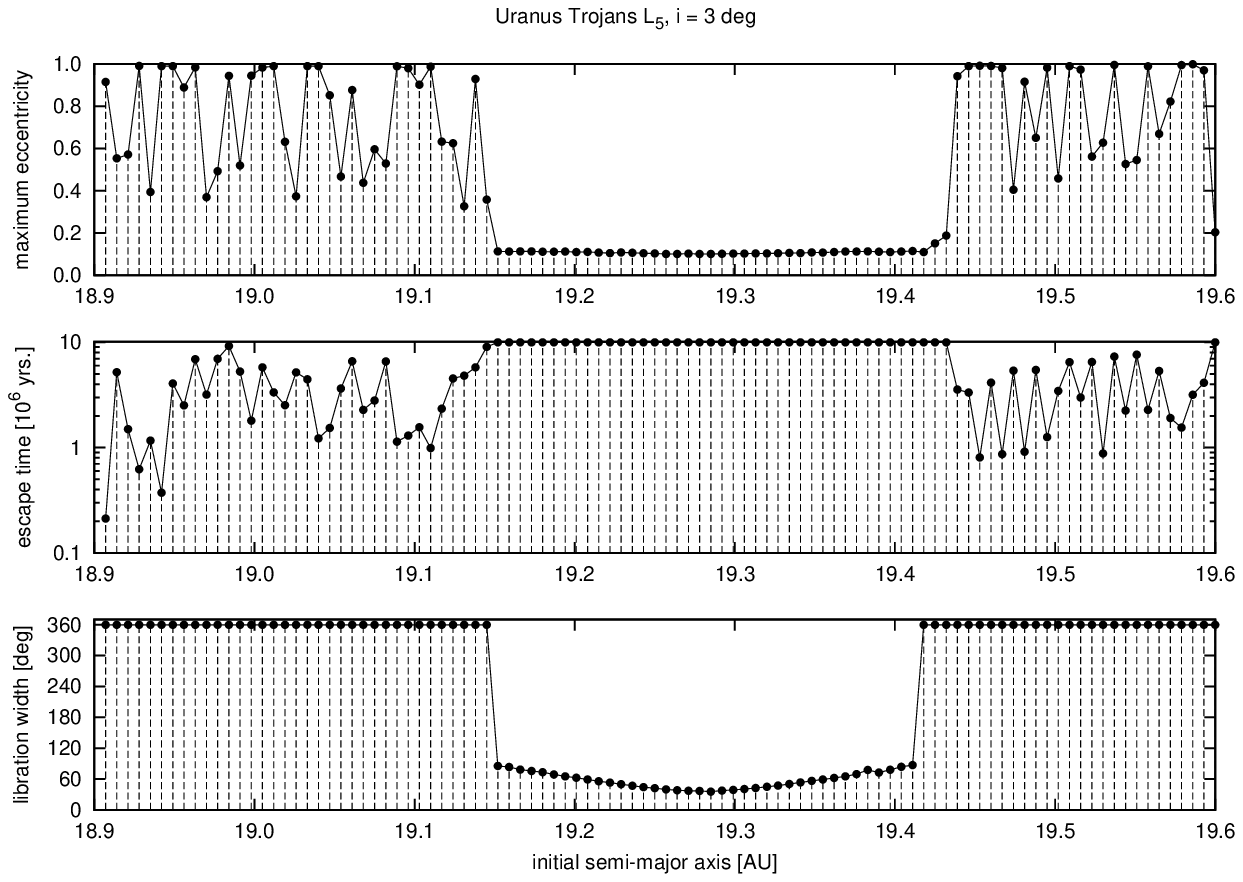}
\caption{{\bf a-cut} for $L_5$: semimajor axis versus the maximum eccentricity
  (top), versus the escape time (middle) and versus the libration width around the
  Lagrange point (bottom).}
\label{fig3}
\end{center}
\end{figure}

As mentioned before these computation are very cumbersome from the point of
view of the CPU times. Therefore we made the following decisions for the
initial conditions of the  {\bf a-cuts}: the inclinations of the Trojans were
-- as mentioned before --
$i=3^\circ$ and the semimajor axes were set to 100 values between $18.9 \le a \le
19.6$ AU, such that the Lagrange points should be in the middle of this interval.
This choice showed the apparent 'asymmetry' between $L_4$ and $L_5$, but
displayed well the actual location of the Lagrange points. This kind of
asymmetry also appeared in the investigations of the Jupiter Trojans
already in the first studies (e.g. \citet{Hol1993, Nes2002}) and was discussed
in more detail in recent papers (\citet{Rob2006, Lho2008}), also for Neptune
(\citet{Zho2009, Dvo2007, Dvo2008}).
The apparent 'asymmetry' is due to the differently chosen initial conditions
for the hypothetical Trojans when one investigates the area around $L_4$ and
$L_5$ because of the presence of Jupiter which changes the location of the
equilibrium points. Strictly speaking they don't even exist in the n-body
problem, but they can be defined as the minimum in the
libration angle. Nevertheless the stable regions are symmetric and it is
sufficient to explore only one (e.g. $L_4$).
In fact comparing the stability diagram of the fictitious Trojans with different
semimajor axes for $L_4$ and $L_5$ first of all the shift in the stable region
is eye-catching: the Lagrange point itself is lying at $a=19.235$ AU, whereas the
middle of the stable area for $L_4$ is at $a=19.18$ AU, and for $L_5$ it is at
$a=19.3$ AU (cf. Figs. \ref{fig2}, \ref{fig3}).
The respective values for position in $a$ for the Lagrange points were
then chosen as the initial semimajor axis
for the {\bf i-cuts} described in the next section.

The time interval to show the extent of the stable region with respect to
the semimajor axis was set to $10^7$ years; we are aware that this is not
enough to determine the effective largeness, but gives a good second estimate
after the results for the $10^6$ years integration for the Lagrange point
$L_4$ of Fig. \ref{fig1}. In fact one can see
that in Fig. \ref{fig2} the outer parts, which still show stable orbits in
the $10^6$ year integration, are unstable, but the central area is still very
stable with librations with amplitudes $A \le 80^\circ $. It has been mentioned
in the former section that on the edge stable and unstable orbits are close to
each other which explains the unstable gap around $a = 19.07$ AU
especially visible in the libration plot (Fig. \ref{fig2} bottom).

This unstable gap is not visible for orbits around the Lagrange point
$L_5$; it shows a smooth stable region with increasing libration amplitudes on
both sides of the equilibrium point itself (Fig. \ref{fig3} bottom).
The different escape times and larger eccentricity values in the
unstable region are not surprising, because these orbits are -- after leaving
the region around the Lagrange point -- in a chaotic state suffering from
multiple encounters with the planets.

\subsection{The {\bf i-cuts}}

\begin{figure}
\begin{center}
\includegraphics[width=4.in,angle=0]{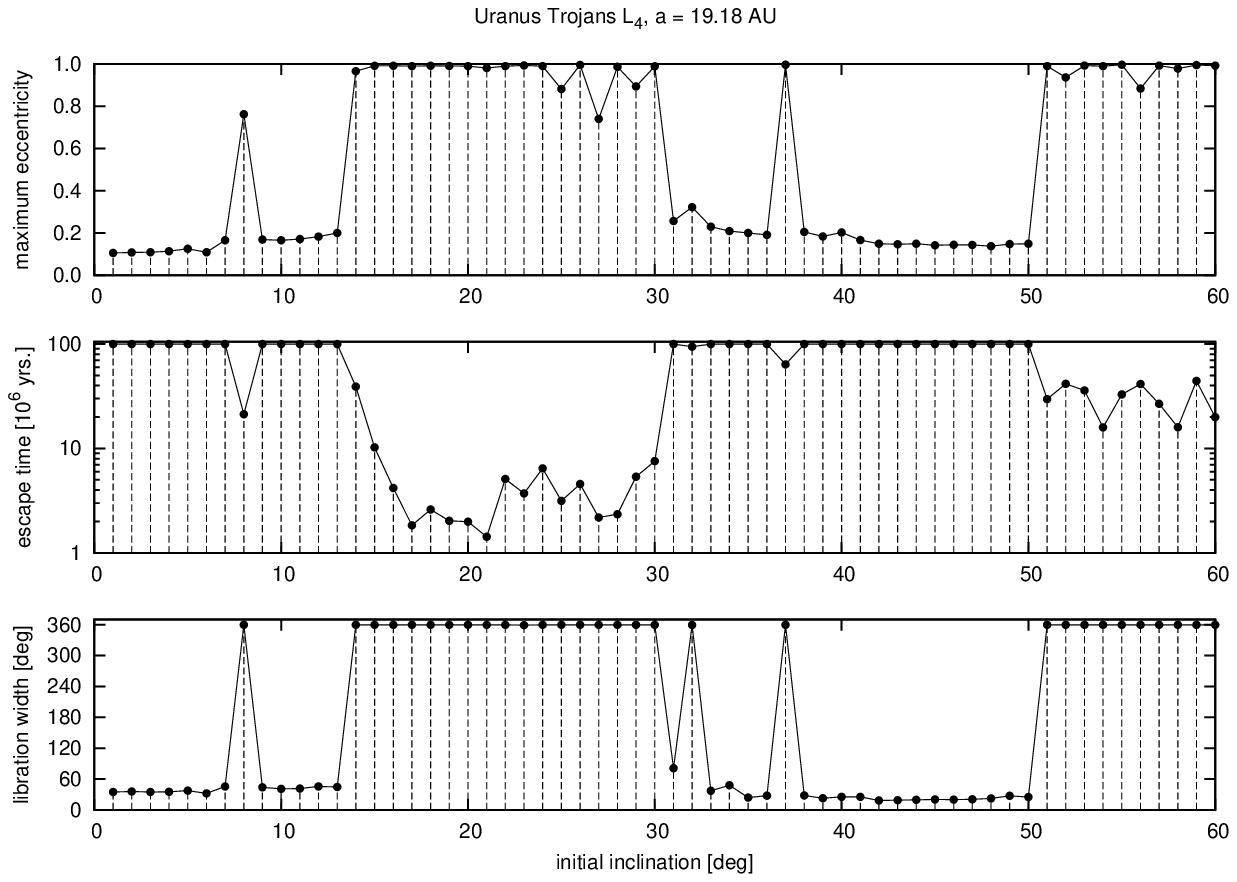}
\caption{{\bf i-cut} for $L_4$: inclination versus the maximum eccentricity
  (top), versus the escape time (middle) and versus the libration width around the
  Lagrange point (bottom).}
\label{fig4}
\end{center}
\end{figure}

\begin{figure}
\begin{center}
\includegraphics[width=4.in,angle=0]{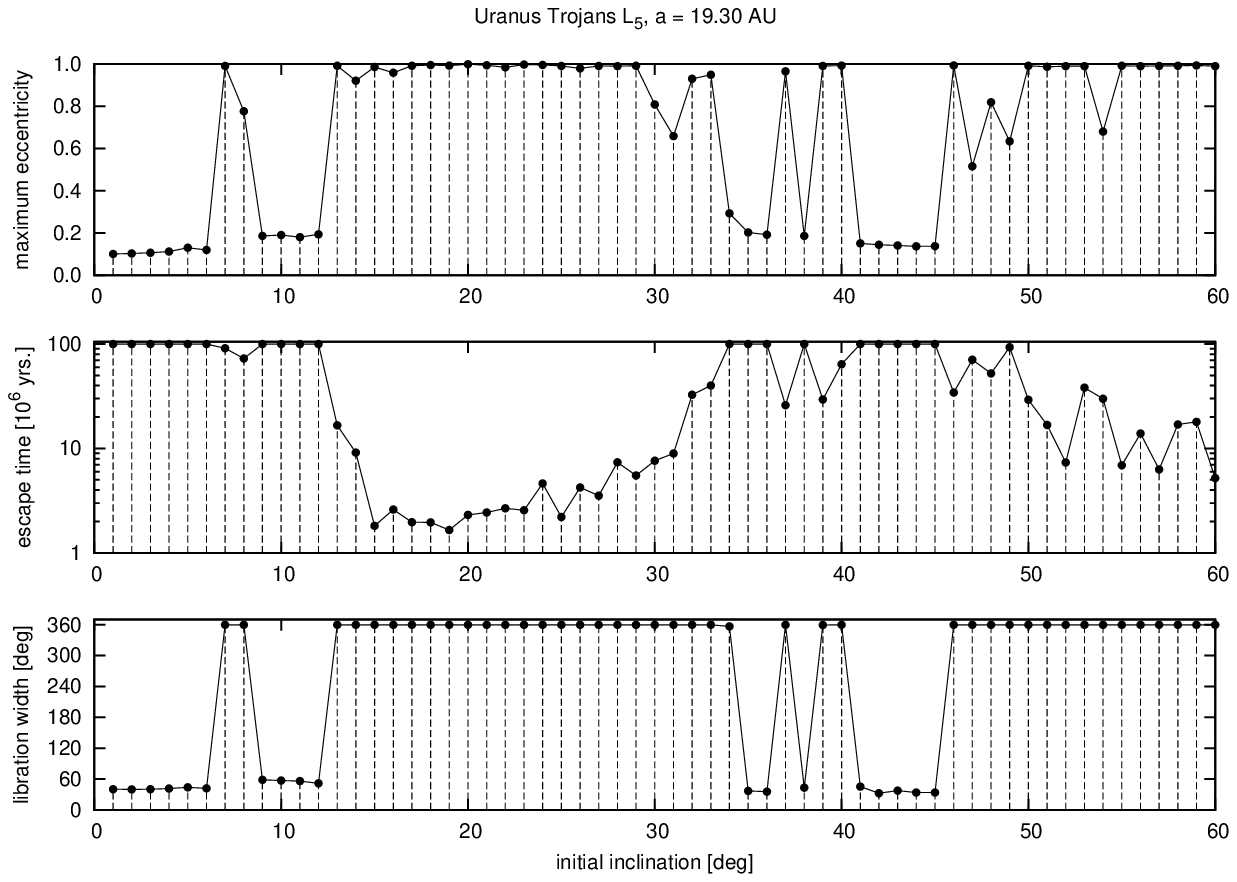}
\caption{{\bf i-cut} for $L_5$: inclination versus the maximum eccentricity
  (top), versus the escape time (middle) and versus the libration width around the
  Lagrange point (bottom).}
\label{fig5}
\end{center}
\end{figure}

In Fig. \ref{fig1} there are two main
regions with stable orbits separated by a curved strip of instability for an
inclination around $i = 16^\circ$. Although we expected that the stable regions will
shrink for an integration time 100 times longer ($10^8$ years), the opening of
an unstable gap around  $i = 8^\circ$ was a surprise. In the next section,
where we carefully discuss the dynamics of three fictitious Trojans, a
possible explanation will be given.

According to our results we can distinguish four different connected region in
Fig. \ref{fig4}, where the orbits are stable with $e_{max} \leq 0.2$ (upper graph),
no escapes occur up to 100 million years (middle graph) and only with small libration angles (bottom
graph):

\begin{description}
  \item{\bf A}  for $0^\circ \le i_{Trojan} \le 7^\circ$
  \item{\bf B}  for $9^\circ \le i_{Trojan} \le 13^\circ$
  \item{\bf C}  for $31^\circ \le i_{Trojan} \le 36^\circ$
  \item{\bf D}  for $38^\circ \le i_{Trojan} \le 50^\circ$
\end{description}

This description and the finding of the regions ${\bf A}$ to ${\bf D}$ has
been done for the $L_4$ environment. The picture for $L_5$ is different in the
following features: the unstable strip at around $i= 8^\circ$ is slightly larger,
also the unstable gap close to $i =  37^\circ$ is broader which makes region
${\bf D}$ significantly smaller. What we expect is
that for even longer integration time the stable region ${\bf C}$ will
disappear and we will be left with three 'islands' of stability with respect
to the inclinations around both Lagrangian equilibrium points. We could not
identify the small instability strip between {\bf A} and {\bf B} with a
special secular frequency (visible also in the signal of the Ura3 orbit). The
instability between {\bf B}, respectively {\bf C} and {\bf D} is caused by
{\bf $g_5$} (Jupiter) and {\bf $g_7$} (Uranus) according to our analysis of the
main frequencies which
corresponds quite well with the
results of \citet{Mar2003} (their Fig. 10). No stable orbits are present for
$i \ge 50^{\circ}$
which is due to the presence of {\bf $g_8$} (Neptune)\footnote{$g_5$ to $g_8$
    are the fundamental frequencies of the giant planets Jupiter (5) to
    Neptune (8)}. A detailed
investigation of all frequencies involved in these regions
is in preparation, where not only cuts
will be used but also the extensions in semimajor axes and inclination.

\subsection{Escapes during 100 Million years}

\begin{figure}
\begin{center}
\includegraphics[width=2.8in,angle=270]{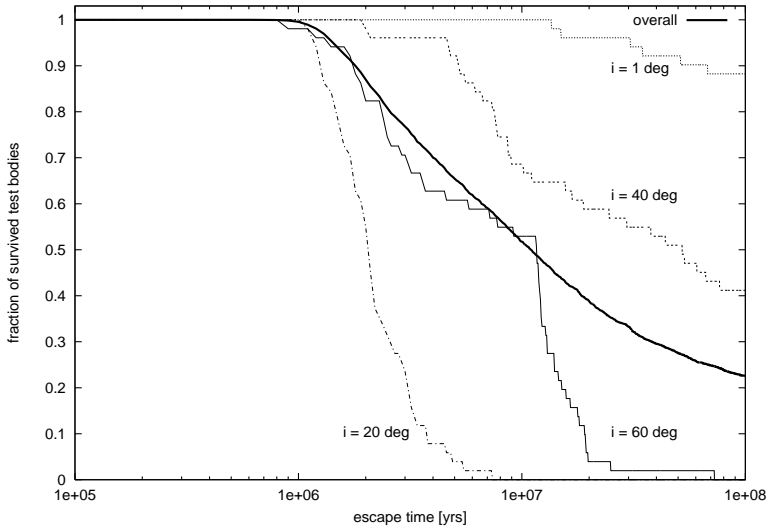}
\caption{Escape times of Uranus Trojans close to $L_4$; the full line
    summarizes the results.}
\label{fig6}
\end{center}
\end{figure}

Another plot was drawn concerning the number of escapers from the region around
$L_4$ of Uranus during 100 million years. Here we show the number of escapers
and compare the results with the ones derived in \citet{Nes2002} (Fig. \ref{fig6}).
In our investigations, where we restricted the initial semimajor axes to
$19.1 \le a \le 19.3$ AU, $\sim 3100$ orbits were integrated.
Here it is possible to distinguish for different inclinations; it is evident
that the escape behaviour is quite different for $i=1,20,40$ and $60^{\circ}$.
For all the inclinations together the plot shows that at the beginning -- the
first million years -- almost no objects escape, but then a constant depletion
up to 100 million years occurs. Finally after the total integration time some
$20\%$ of the original population survive, which is comparable to the results
derived by \citet{Nes2002}.
The different escaping rate for different initial inclinations reveals that
there are different mechanisms taking effect on different timescales, which
will be analyzed in detail in our future work.

\section{Long term integrations}

To assure that the determined stable regions will survive for times up to the age of
the solar system three orbits close to the Lagrange point $L_4$ were integrated
for $5 \times 10^9$ years for low inclined orbits. The initial
conditions were the
ones of Uranus except
for the inclinations:

\begin{itemize}
\item {\bf Ura1} with $i_{Trojan}=i_{Uranus}$,
\item {\bf Ura2} with $i_{Trojan}=i_{Uranus}+2^\circ$,
\item {\bf Ura3} with $i_{Trojan}=i_{Uranus}+4^\circ$.
\end{itemize}

\begin{figure}
\begin{center}
\includegraphics[width=3.0in,angle=270]{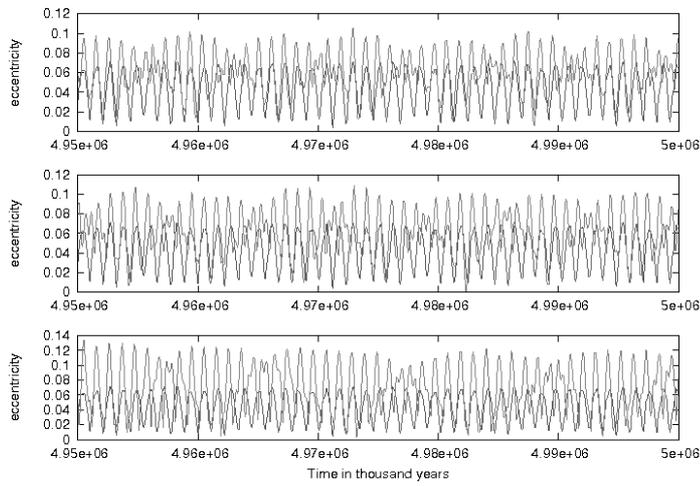}
\caption{The eccentricities of 3 fictitious Uranus Trojans after 5 billion
  years compared to the one of Uranus: Ura1(top), Ura2 (middle) and  Ura3 (bottom).}
\label{fig7}
\end{center}
\end{figure}

\begin{figure}
\begin{center}
\includegraphics[width=3.0in,angle=270]{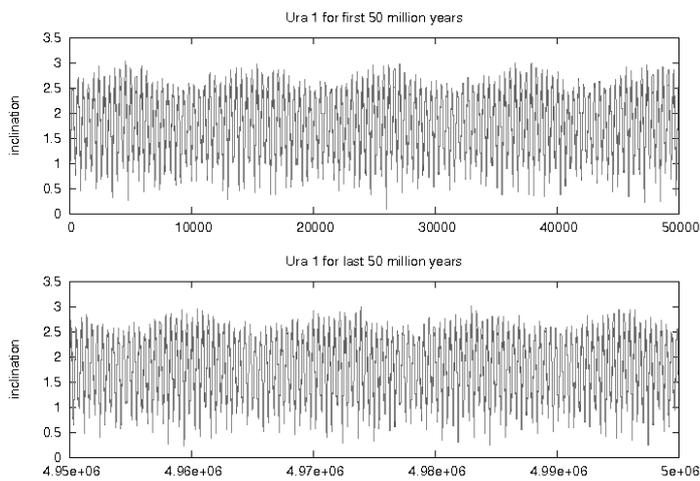}
\caption{Inclination of Uranus and Ura1 for the first 50 and the
  last 50 million years of integration (upper, respectively lower
  graph). Note that the signal for Ura1 and Uranus are undistinguishable.}
\label{fig8}
\end{center}
\end{figure}

\begin{figure}
\begin{center}
\includegraphics[width=3.0in,angle=270]{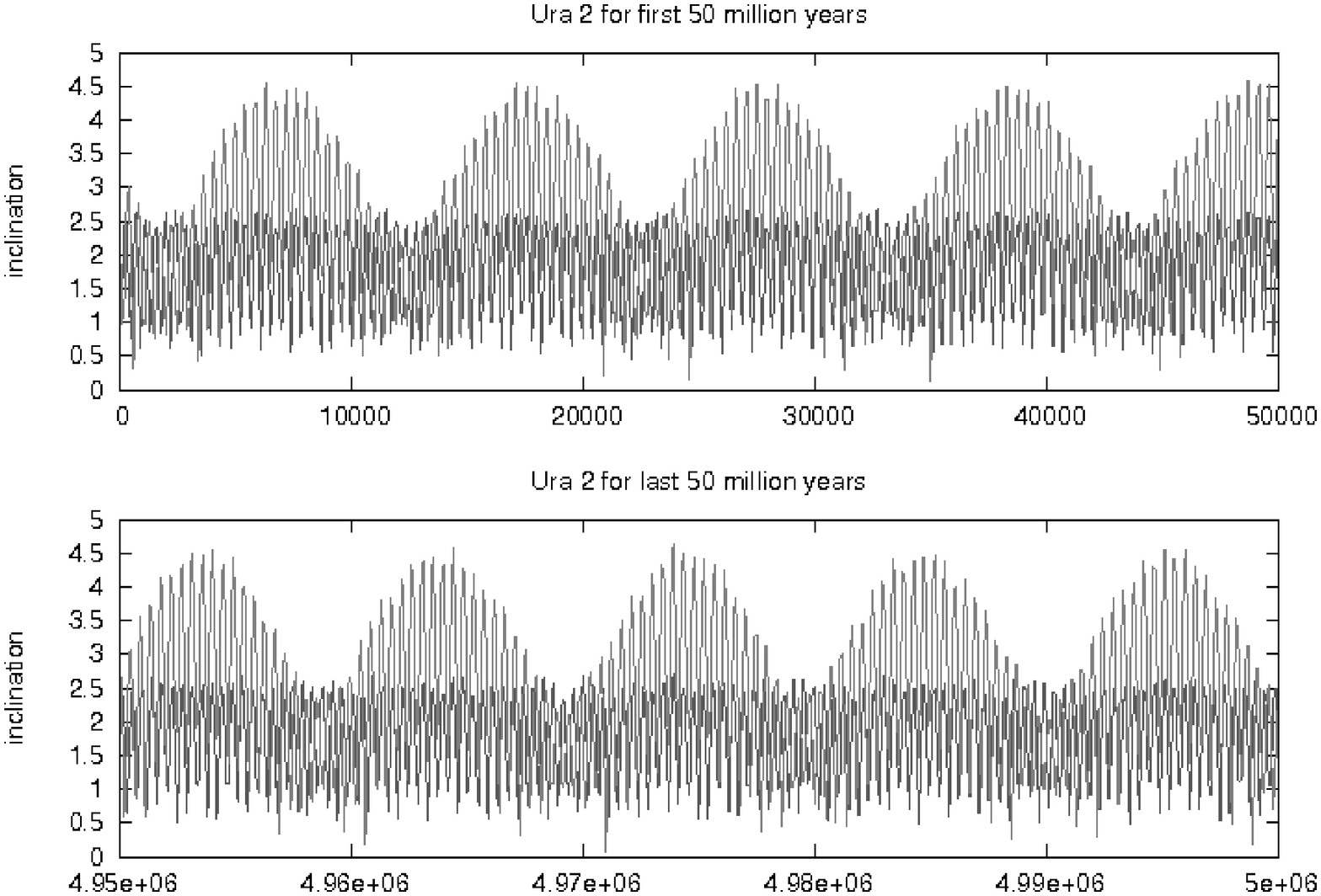}
\caption{Inclination of Uranus and Ura2 for the first 50 and the last 50 million
  years of integration.}
\label{fig9}
\end{center}
\end{figure}

\begin{figure}
\begin{center}
\includegraphics[width=3.0in,angle=270]{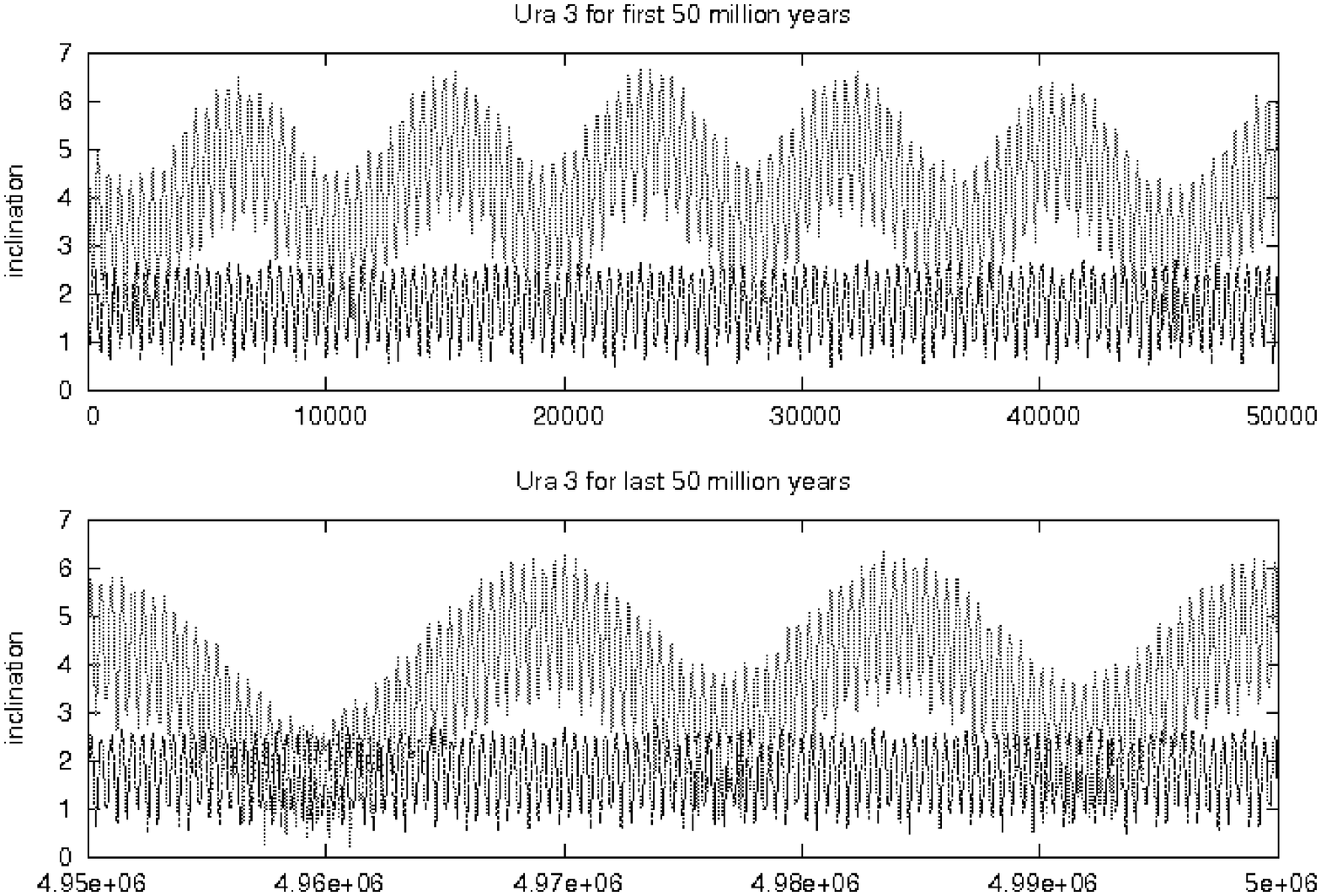}
\caption{Inclination of Uranus and Ura3 for the first 50 and the last 50 million years of integration.}
\label{fig10}
\end{center}
\end{figure}

In the respective plot where we compare the eccentricities of
fictitious Trojans directly with Uranus (Fig. \ref{fig7}), it is
evident that there is not a big difference in the time development of the
eccentricities; only the most inclined Trojan
{\bf Ura3} suffers from larger amplitudes than {\bf Ura1} and {\bf Ura2} in $e$. The comparison
of the inclination plots unveils why there is an instability gap at about
$i=6^\circ$: Whereas on the first graph (Fig. \ref{fig8}) no difference is visible at all between
the Trojan {\bf Ura1} and Uranus itself, already in the graph for {\bf Ura2}
(Fig. \ref{fig9}) a significant feature can be observed.
Large periodic changes occur with $i_{max} \geq 4^\circ$. From the graph of {\bf  Ura3}
(Fig. \ref{fig10}) one can see that the period of the variation of the inclination at the
end of the 5 billion years is only half of the period at the beginning.

\begin{figure}
\begin{center}
\includegraphics[width=4.2in,angle=0]{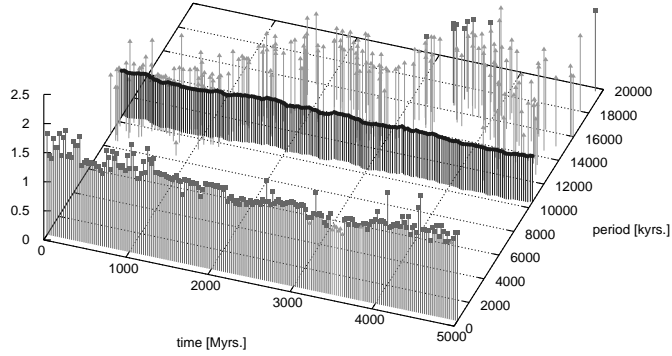}
\caption{The two main periods in the element $k= e \cos(\Omega+\omega)$ for
the 5 billion years integration of {\bf  Ura3} namely the $4.4 \times 10^5$
years period  and the
period which changes between $\sim 10^7$ and   $2 \times 10^7$ years compared to
the 'same' period of   $ \sim 10^7$ of {\bf  Ura2}  which is shown as as
straight {\sc wall} in the middle of the graph: periods (y-axis) are plotted versus
the time (x-axis) and the amplitudes (z-axis).}
\label{fig11}
\end{center}
\end{figure}

A carefully undertaken frequency analysis showed that in contrast to  {\bf
  Ura1} and {\bf Ura2}
for  {\bf  Ura3} a dramatic change in the
main frequencies occured during the dynamical evolution.
This is a sign of chaotic behaviour as was already
pointed out in many papers (e.g \citet{Las1990, Las1994, Las1997}).
To visualize this change we used the
technique of running windows, each section spanning 50 million years with an overlap
of 25 million years. The changes are well visible for the larger periods of
some 12 to 18 million years which also show significantly larger amplitudes
than the second largest period of about $4.4 \times 10^5$ years (Fig. \ref{fig11}).

\begin{figure}
\begin{center}
\includegraphics[width=3.0in,angle=270]{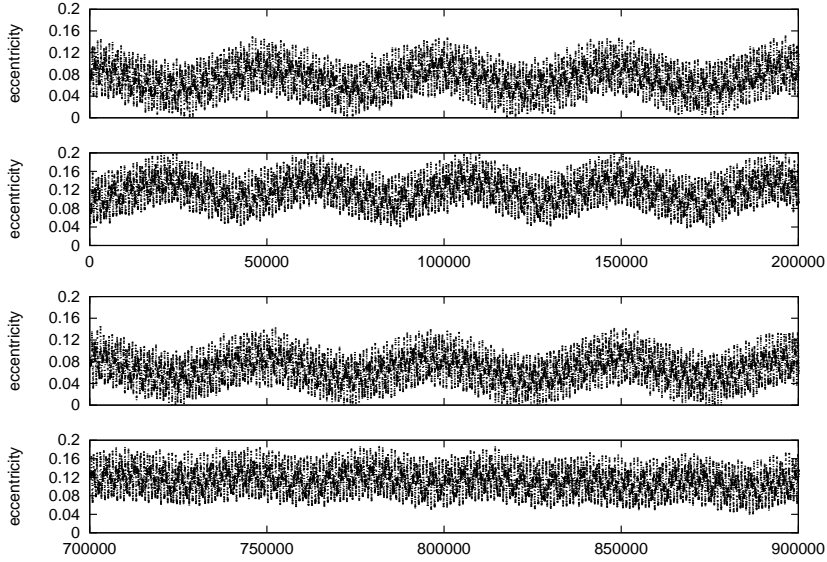}
\caption{Eccentricity of two orbits close to the lower instability border for
  inclinations $i=42^{\circ}$ ($1^{st}$ and $3^{rd}$ panel)
  and for $i=40^{\circ}$ ($2^{nd}$ and $4^{th}$ panel)
  for the first 200 Myrs (upper panels) and
  from 700 Myrs to 900 Myrs (lower panels); time is measured in Myrs.}
\label{fig12}
\end{center}
\end{figure}

In addition, at the lower end of the high inclination window, we made numerical
integrations up to 1Gyr which showed that even orbits close to the unstability
window have the typical behaviour of changing the periodicity. This can be
seen already after several hundred million years in Fig.\ref{fig12} for two orbits,
one is well inside the stability region ($i=42^{\circ}$) and the other one already
close to the edge ($i=40^{\circ}$). In fact the last one becomes unstable
after another 100 Myrs (at 946 Myrs); this behaviour gives notice already
when one compares the dynamics for the first 200 Myrs and the last 200 Myrs of
this hypothetical Trojan (compare the $2^{nd}$ and $4^{th}$ panel in Fig. \ref{fig12}).

\section{Conclusions}

In our study we explored the regions of possible Trojan asteroids of
Uranus for a large range of their inclinations up to $i= 60^\circ$,
which has never been done before. The method of determining these
regions was the analysis
of the results of straightforward numerical integrations of the full
equations of motion in the model of the OSS. Besides the escape from the region we also checked the
amplitudes of libration of these fictitious asteroids around the Lagrangian
equilibrium points. To assure the survival of the 'stable' Trojans we did also
integrations of three specially chosen Trojans on low inclined orbits for the
age of the Solar system of $4.5 \times 10^9$ years, where the
accurate analysis of the frequencies involved was undertaken.
Our results showed that close to the first instability window
$6^\circ \leq i_{Trojan} \leq 8^\circ$ the strong variation of the frequency of
a special secular resonance  indicates chaotic motion. Two other unstable
windows with respect to the inclination are $14^\circ \leq i_{Trojan} \leq
31^\circ$ and around $i_{Trojan} = 37^\circ$. From $i_{Trojan} \geq 50^\circ$
(for $L_4$) and  $i_{Trojan} \geq 45^\circ$ ($L_5$) no stable Trojan orbits
exist for Uranus. These stable islands in inclination are surrounded by more or
less small stable regions in semimajor axis.

According to the results of \citet{Mor2005} Jupiter Trojans could be trapped
during the phase of early migration of the planets in our Solar system. If
that was also the case for Neptune\footnote{up to now 6 Neptune Trojans were
observed} and Uranus Trojans (and why not) we should be able to detect them.
So the question is: why don't we observe them?

\section{Acknowledgements}

For the realization of this study {\'A}. Bazs{\'o} needs to thank the Austrian Science Foundation (FWF) project
18930-N16. L.-Y. Zhou thanks the support from Natural Science
Foundation of China (No. 10403004, 10833001) and the National Basic
Research Program of China (2007CB814800).


%
%
%

\end{document}